\documentclass[sigconf, nonacm, screen=true, review=false, anonymous=false, authordraft=False]{acmart}

\usepackage{tabularx}
\usepackage{dcolumn} 
\newcolumntype{d}[1]{D{.}{.}{#1}}

\usepackage{subcaption}

\usepackage{todonotes}
\let\xtodo\todo
\renewcommand{\todo}[1]{\xtodo[inline,color=green!50]{#1}}

\AtBeginDocument{%
  }

\setcopyright{CC}
\setcctype{by-sa}

\begin{document}

\title[How Can Mixed Reality Benefit From Physiologically-Adaptive Systems?]{How Can Mixed Reality Benefit From Physiologically-Adaptive Systems? Challenges and Opportunities for Human Factors Applications}


\settopmatter{authorsperrow=3}

\author{Francesco Chiossi}
\orcid{0000-0003-2987-7634}
\affiliation{
  \institution{LMU Munich}
  \country{Germany}
}
\email{francesco.chiossi@um.ifi.lmu.de}

\author{Sven Mayer}
\orcid{0000-0001-5462-8782}
\affiliation{%
  \institution{LMU Munich}
  \postcode{80337}
  \country{Germany}}
\email{info@sven-mayer.com}

\renewcommand{\shortauthors}{Chiossi and Mayer}


\begin{teaserfigure}
    \centering
    \includegraphics[width=\linewidth]{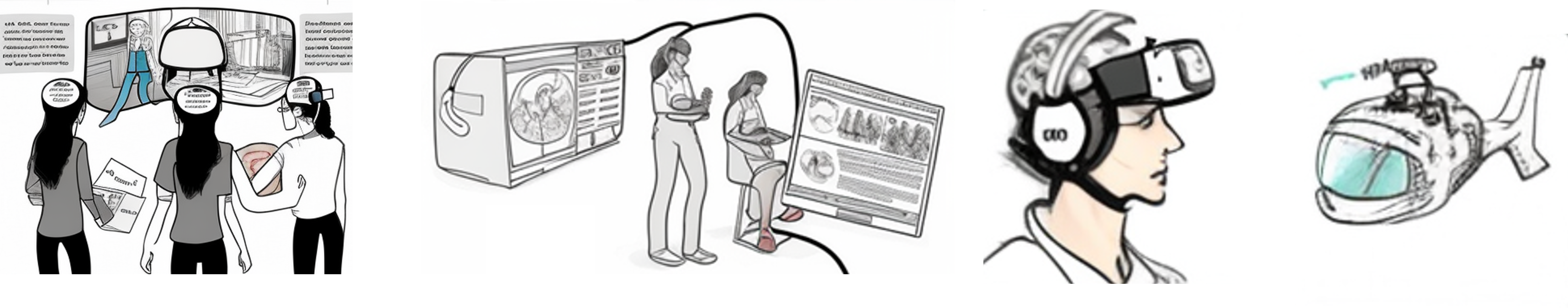}
    \caption{This paper explores the limitations of current Mixed Reality (MR) experiences and how physiologically-adaptive systems can optimize interactions and enhance user experiences in various domains, such as healthcare, education, and entertainment. Physiologically-adaptive systems have the potential benefit of providing users with more personalized and engaging experiences. However,  addressing ethical and privacy concerns is a fundamental issue in the HCI community when dealing with implicit systems and inputs over which users have limited control.}
    \Description{The picture  shows a person wearing a mixed reality headset, standing in front of a large screen displaying virtual content. Next, we have two people in MR in healthcare, one is  a practitioner, and the other is a patient. Finally, we have a training session in MR with a user looking at a helicopter in MR.}
    \label{fig:teaser}
\end{teaserfigure}

\begin{abstract}
Mixed Reality (MR) allows users to interact with digital objects in a physical environment, but several limitations have hampered widespread adoption. Physiologically adaptive systems detecting user's states can drive interaction and address these limitations. Here, we highlight potential usability and interaction limitations in MR and how physiologically adaptive systems can benefit MR experiences and applications. We specifically address potential applications for human factors and operational settings such as healthcare, education, and entertainment. We further discuss benefits and applications in light of ethical and privacy concerns. The use of physiologically adaptive systems in MR has the potential to revolutionize human-computer interactions and provide users with a more personalized and engaging experience.
\end{abstract}



\maketitle

\section{Introduction}

Mixed reality (MR) systems encompass a broad spectrum that spans from physical reality to virtual reality (VR), including instances that involve overlaying virtual content over physical one, i.e., Augmented Reality (AR), as well as those that use physical content to enhance the realism of virtual environments, i.e. Augmented Virtuality (AV) \cite{speicher2019}. These instances are typically predefined for a seamless physical and virtual content blend.

MR enables users to interact with digital objects in a physical environment, resulting in immersive and engaging experiences. However, several limitations have hampered the widespread adoption of MR technology \cite{skarbez2021revisiting, milgram1994taxonomy}. In recent years, researchers have begun to investigate the use of physiologically adaptive systems to address these limitations by developing systems that can respond in real-time to the user's physiological state \cite{baker2022adaptive}.

Physiologically adaptive systems belong to a group of adaptive systems that employ physiological signals to generate personalized and captivating experiences. They are based on user's physiological signals as a form of input, such as peripheral measures,  e.g., electrocardiogram \cite{munoz2020psychophysiological} or electrodermal activity \cite{chiossi2022virtual}, and central physiological measures, such as electroencephalography (EEG) \cite{tremmel2019estimating, chiossi2023exploring}, and Functional near-infrared spectroscopy (fNIRS) \cite{afergan2014} produce real-time feedback and responses based on the user's physiological state. Physiologically-adaptive systems are based on classic control theory \cite{wiener2019cybernetics}. This theory involves three main steps: physiological data acquisition and processing, transformation into a system response, and shaping the expected psychophysiological response from the user. These so-called "Biocybernetic control loops" \cite{fairclough2012construction,chiossi2022it} employ a negative control to detect deviations from the optimal state and prompt changes in the system to encourage a desirable user's state. This process is crucial in creating a responsive and personalized experience for the user.

Considering that physical and virtual reality are the two extremes of the MR continuum, this provides a favourable setting for developing adaptive systems. Adaptive systems can tailor the MR experience to the user's needs and goals by leveraging this continuum, assisting them in achieving optimal performance \cite{cote2009cognitive}, immersion \cite{grassini2020questionnaire}, and engagement \cite{doherty2018}.

This paper aims to investigate the potential applications of physiologically adaptive systems in MR and discuss their advantages and disadvantages. We will specifically look at the benefits of physiologically adaptive systems in addressing the limitations of MR technology and discuss their potential applications.

First, we review the definition of MR and its various forms. We will also look at the current limitations of MR technology and the issues that must be addressed to improve its usability and effectiveness. We will then define physiologically adaptive systems and discuss their characteristics and potential benefits.

Second, we discuss potential applications of physiologically adaptive systems in human factors and applied MR settings. For example, healthcare professionals can use such systems to create more engaging and effective patient therapies by providing real-time feedback and support based on the patient's physiological state. By adapting to the student's cognitive and physiological state, these systems can be used in education to create more immersive and engaging learning experiences. By adapting to the player's physiological state and creating more personalized and engaging experiences, these systems can be used in entertainment to create more engaging and immersive games and simulations.

Finally, we highlight challenges for physiologically adaptive systems in MR, including technical and theoretical constraints and ethical and privacy concerns. We discuss potential solutions and strategies for dealing with such fundamental issues.

\section{Mixed Reality}

The predominant definition of MR is the one provided in the seminal work by Milgram and Kishino \cite{milgram1994taxonomy}, referring to the merging of real and virtual worlds in a seamless and interactive environment. It is an interaction spectrum that blends physical and digital realities to create a new, immersive experience for the user. 

Recently, this perspective has been reviewed by Skarbez et al. \cite{skarbez2021revisiting}. Their revised taxonomy consists of three dimensions: immersion, coherence, and extent of world knowledge. Immersion is determined by a system's objective hardware device specifications and is related to the feeling of spatial presence experienced by the user \cite{slater2009place}. Coherence refers to the conformity of different sensory information perceived during an XR experience, leading to an increased plausibility illusion of the experience \cite{skarbez2017survey}. The extent of world knowledge describes the degree of reality incorporated into an MR experience, influencing the user's real-world awareness \cite{milgram1995augmented}. The authors focus on immersion and coherence and consider important environmental cues that influence the extent of world knowledge.

Latoschik and Wienrich \cite{latoschik2022congruence} provide a third perspective that emphasizes that congruence activations between cognitive, perceptual, and sensory layers contribute to MR plausibility. The authors argue that device specifications, like the field of view or resolution, impact device-specific sensory congruence, while content transparency affects congruence. These congruences ultimately affect the plausible generation of spatial cues and spatial presence.

\subsection{Current Limitations for MR Systems Adoption}
Despite technical and design advancements in Mixed Reality (MR) technology, significant limitations still prevent it from reaching its full potential and adoption by the general public and professionals. Now, we highlight four main factors that contribute to such limitations. 

First, a limited field of view (FoV) represents an initial issue in many MR systems. FoV is the area that the user can see through the display, and it is often constrained by the physical size of the device's screen or lenses \cite{gagnon2021gap}. A limited FoV can reduce immersion and realism and lead to visual discomfort \cite{ragan2015effects}, especially when the user must frequently turn their head to view the content \cite{ren2016evaluating}.

Secondly, we identify limited interactivity as a primary constraint for MR adoption. MR systems often rely on gesture recognition or voice commands \cite{handosa2018extending}, which can be imprecise and unreliable, leading to frustration and reduced user engagement. This limitation can be a significant barrier to adopting MR in some domains, such as entertainment applications \cite{stapleton2002applying}, i.e., gaming or when this adds up to an existing cognitive load, such as in education settings \cite{shaytura2021}.

Third, while modern MR devices can display highly detailed virtual content alone, their embedding into physical reality hinders the efficiency of their plausibility \cite{min2019}, ultimately leading to reduced realism. On the contrary, high levels of realism can strengthen the efficiency of training simulations \cite{goncalves2023}. Still, on the other side, when increasing details and amount of virtual content, we implicitly impact the MR visual complexity \cite{olivia2004identifying} that has been shown to influence behavioural performance and physiological arousal  \cite{makransky2019adding, pierno2005effects, ragan2015effects}.

Finally, limited adaptability is another significant limitation of MR systems. Many MR applications are pre-defined and cannot adapt to the user's changing needs or physical state. This limitation can reduce the effectiveness of MR applications and lead to reduced user engagement and long-term usage.

\begin{figure*}
  \centering
  \includegraphics[width=\textwidth]{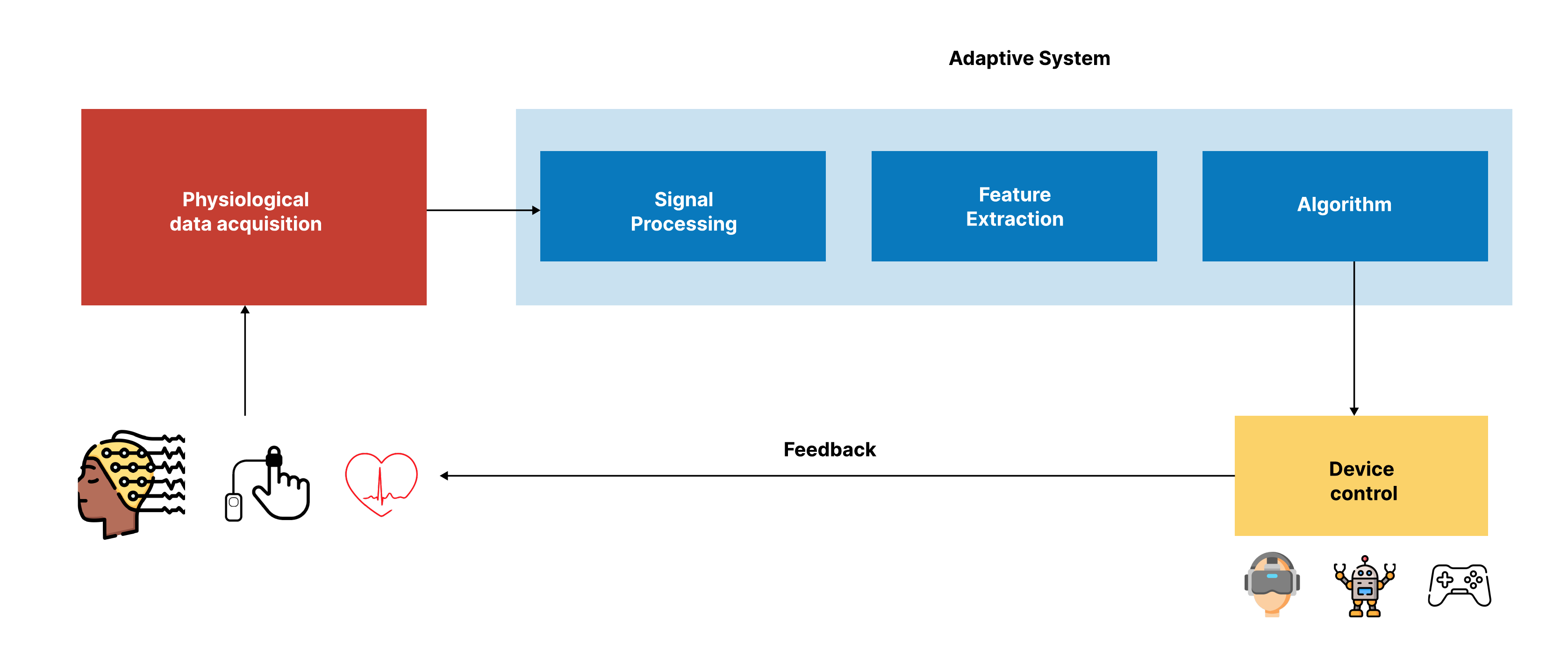}
  \caption{In a biocybernetic control loop, the Adaptive System continuously process, and extract informative features from the physiological signals and detect the user state based on an algorithm. Thus, it adjusts its behaviour for device control for diverse applications and provides returns feedback to the user. This closed-loop allows for iterative adaptations and optimization of visualization, content and interaction.}
  \label{fig:example}
\end{figure*}

\section{Physiologically-Adaptive Systems in MR}
Physiologically-adaptive systems are systems designed to interact with and respond to the physiological states and changes of the human body. These systems typically employ sensors and algorithms to monitor and analyze physiological signals such as ECG, EDA and EEG to drive interactions towards a specific state based on the cybernetics approach \cite{tzafestas2022systems}.
The cybernetics approach found various applications ranging from developing new control channels \cite{monkaresi2012} to task adaptation in response to changes in workload \cite{prinzel2000closed} and motivation \cite{ewing2016evaluation}. 

However, most of the work focused on desktop settings. Only recently, MR settings are proliferating and enabling the creation of environments and interactions far more engaging and expressive than traditional desktop programs \cite{putze2020brain, gramann2021grand}.

MR is now one of the most favourable environments for physiological computing systems. MR enables online adjustments and adaption of visualizations, digital content, blending, and interactions that resemble real-world ones. However, it is not currently feasible in physical settings (VR) or augmenting them (AR, AV). Introducing physiological interaction into MR can increase its ability to monitor and adapt to implicit human behaviour. Physiologically adaptable MR systems can identify user states and direct interaction characteristics toward a (shared) objective depending on physiological input.

\subsection{Benefits of Physiologically-Adaptive systems for MR}

With regard to the limitations of MR adoption, we identify how physiologically-adaptive systems can enhance MR interaction and address possible usability constraints. 

While the limited field of view (FoV) in MR devices is primarily a hardware limitation that may be challenging to address through physiological adaptivity alone, monitoring attention and gaze can still play a role in enhancing the user experience within the existing FoV limitations. 
Physiological inputs such as eye gaze and torso movements and their temporal alignment can be employed for attention, interest, and intent detection and as context and input \cite{sidenmark2019eye}. Moreover, EEG features such as alpha and theta oscillations discriminated between internally and externally directed attention in AR \cite{vortmann2019eeg} and VR settings \cite{magosso2019eeg}. This information can be used to dynamically adjust the field of view of the MR system, for example, by zooming in on areas of interest, providing multisensory cues to direct attention towards hidden areas, and blurring distracting information.

Limited interactivity refers to situations where the user has limited ability to control or manipulate the virtual objects in the MR environment. This can occur due to factors such as the complexity of the interface or the user's cognitive workload. 

Limited interactivity can benefit from neuro-, and electrophysiological measures such as EEG and fNIRS for workload \cite{de2014fast, hirshfield2018}, and attention detection \cite{souza2021attention}, enlarging the design space for interaction. For instance, if the user is experiencing cognitive overload or boredom  \cite{ewing2016evaluation}, the system can simplify the interface or adjust the task difficulty level to maintain engagement and interest. Additionally, if the user is experiencing unpleasant states such as frustration or anxiety \cite{luong2021survey}, the system can distract the user with positive stimuli to distract from their emotional state \cite{stamp2022influence} and maintain their attention on the task \cite{fairclough2023functional}.

Third, the limited realism could be controlled and adapted based on autonomic arousal, i.e., EDA or ECG, for leveraging its effect on the user's physiological activation. This physiological input can be used to adjust the level of MR visual fidelity, for example, by adding or removing sensory cues to enhance the user's emotional experience \cite{chen2017exploring, weiwei2022}, or support target detection, when engaged in visual search \cite{chen2018investigating, hancock2013improving}.

Lastly, physiologically-adaptive systems are central to increasing reactivity and adaptability. Employing physiological data as a passive input and concurrently adapting either task or environmental features can allow for more dynamic interaction, controlling for undesirable states such as anxiety or boredom \cite{pekrun2010boredom, baker2022adaptive}, improve motivational engagement \cite{ewing2016evaluation}, and therefore allowing users to maintain focus on the current task and perform optimally.

\subsection{Potential Applications of Physiologically Adaptive Systems in Applied MR Settings}
This combination of implicit physiological monitoring and MR environment adaptation can be defined as a closed-loop model. Since their original conception and design in the seminal work of Pope et al. \cite{pope1995biocybernetic}, biocybernetic closed loops have had many implications in human factors, and applied settings, such as aviation \cite{haarmann2009combining}, healthcare \cite{munoz2018closing}, and other high-demanding environments \cite{prinzel2000closed}.
 
We envision three operational settings where physiologically-adaptive MR environments can be profitable: healthcare, education, and training. 

Physiologically adaptive systems in the healthcare industry can deliver customized therapies suited to the patient's psychophysiological condition. Physiological measures, for example, can be used in mental health to assess physiological signals related to stress, anxiety, and depression. Such information improves the patient's exposure therapy, leveraging the degree of realism or intensity of the phobic stimuli presented either in VR or in AR \cite{kamkuimo2020dynamic}. Similarly, adaptive systems may be used in physical therapy to monitor patients' progress and offer real-time feedback on their movements, allowing therapists to change the intensity of exercises to guarantee optimal recovery and rehabilitation \cite{badesa2016, dietz2022}.

In educational settings,  physiologically adaptive systems can be used to improve learning outcomes. Recently, many companies and educational institutions have allocated considerable resources to transitioning from traditional desktop education to immersive MR applications, expecting that a higher level of immersion would correspond to increased motivation and learning. Physiological monitoring can aid in technology-based educational decision-making to assist cognitive, i.e., information processing \cite{yuksel2016},  emotions,i.e., frustration \cite{harley2017developing}, and motivation and metacognitive \cite{ewing2016evaluation}, i.e., self-regulation behaviours of learners \cite{harley2015multi}. Related to educational settings are also the professional training MR environments \cite{zahabi2020adaptive}.

Finally, the entertainment industry can benefit from the design of physiologically-adaptive games \cite{munoz2017biocybernetic}. Besides adjusting the game realism to support immersion \cite{hudlicka2009} or employing dynamic difficulty adjustments \cite{moschovitis2022}, adaptive gaming can pursue and drive interactions towards less socially acceptable goals. For example, Moschovitis and Denisova \cite{moschovitis2022} showed how they could increase game engagement using a biofeedback-controlled game that elicited physiological responses associated with fear and anxiety. Their results show how stimuli perceived as unpleasant on the surface might result in a positive subjective outcome. Finally, gamification approaches can benefit entertainment purposes and be applied and generalized to different settings, such as therapy, treatment of anxiety and cognitive rehabilitation and training.

\section{Ethical and Privacy Considerations for Implementing Physiologically Adaptive Systems in Mixed Reality}

Within our perspective endorsing a progressive implementation and investigation for physiologically adaptive systems in MR, we have to foresee downsizes and considerations regarding ethics and privacy. 

One of the primary ethical considerations for systems that employ data over which users do not have complete explicit control is the issue of informed consent. Users must be fully aware of how physiological data are collected, used, and shared. This is relevant when their data are employed for model training and validation. 

Secondly, physiological states can underlie different emotional valences, implying that such systems might manipulate or influence users' emotions. Therefore, researchers must prioritize ethical design and inform participants about which state the system is optimizing for. Lastly, they should allow participants to return to a neutral affective state if users perceive their final state as undesirable. This is critical as users must retain control over the adaptation and state adjustment process.

Third, privacy concerns are associated with physiologically adaptive systems in MR. This perspective was already raised by Fairclough \cite{fairclough2008}, highlighting how symmetrical interaction and adaptation between systems and users might lead to asymmetrical data usage and protection. Again, Hancock and Szalma \cite{hancock2003future} highlight that if a physiological computing system respects data protection rights, individuals should retain formal and legal ownership of their psychophysiological data. This implies that any third party should receive access to such information only with approval by the user. This is relevant considering that physiological data might not only underlie specific cognitive or affective states and be used for medical diagnostic purposes.

An initial compromise solution is using a privacy-by-design approach by embedding privacy considerations into every stage of the design and development process. This includes conducting privacy impact assessments, implementing privacy-enhancing technologies, and using privacy-preserving data collection in every implementation stage of the physiologically-adaptive systems.

\section{Conclusion}
In conclusion, MR technology holds great potential for creating immersive and engaging experiences, especially when employing physiologically adaptive systems that allow users to interact with personalized visualizations, contents and interactions. We highlighted how MR experiences could overcome challenges and limitations by embedding biocybernetic paradigms in their systems and depicted future concerns for their implementation. HCI, MR, and adaptive systems research fields can all benefit from the enormous potential of adopting and exploring physiological computing and interaction paradigms. However, such opportunities will only be realized if these fundamental difficulties are addressed by present research in this area.

\begin{acks}
Francesco Chiossi was supported by the Deutsche Forschungsgemeinschaft (DFG, German Research Foundation), Project ID 251654672 TRR 161.
\end{acks}

\bibliographystyle{ACM-Reference-Format}
\bibliography{main}


\end{document}